\documentclass[openacc]{rsproca_new}

\begin{document}

\title{Paradox resolved: The allometric scaling of cancer risk across species}

\author{
Christopher P. Kempes$^{1}$, Geoffrey B. West$^{1}$ and John W. Pepper$^{3}$}

\address{$^{1}$The Santa Fe Institute, Santa Fe, NM 87501, USA\\
$^{2}$National Cancer Institute, Division of Cancer Prevention, Bethesda, MD 20814, USA}

\subject{Theoretical Biology, Physiology, Biophysics}


\corres{Christopher P. Kempes\\
\email{ckempes@santafe.edu}}

\begin{abstract}
Understanding the cross-species behavior of cancer is important for uncovering fundamental mechanisms of carcinogenesis, and for translating results of model systems between species. One of the most famous interspecific considerations of cancer is Peto's paradox, which asserts that organisms with vastly different body mass are expected to have a vastly different degree of cancer risk, a pattern that is not observed empirically. Here we show that this observation is not a paradox at all but follows naturally from the interspecific scaling of metabolic rates across mammalian body size. We connect metabolic allometry to evolutionary models of cancer development in tissues to show that waiting time to cancer scales with body mass in a similar way as normal organism lifespan does. Thus, the expectation across mammals is that lifetime cancer risk is invariant with body mass. Peto's observation is therefore not a paradox, but the natural expectation from interspecific scaling of metabolism and physiology. These allometric patterns have theoretical implications for understanding life span evolution, and practical implications for using smaller animals as model systems for human cancer. 
\end{abstract}


\begin{fmtext}

\end{fmtext}


\maketitle

\section{Introduction}

Cancer is not confined to humans, but occurs throughout the animal kingdom, and perhaps beyond \cite{aktipis2015cancer}. This opens possibilities for deeper understanding of cancer through comparative studies and comparative oncology\cite{paoloni2007comparative, kitsoulis2020occurrence}. The evolutionary thinking driven by species comparisons has been employed previously in understanding tumor growth and metastasis, and is also informative regarding cancer risk. The comparative approach thus has much to contribute to understanding cancer epidemiology and to advancing cancer prevention \cite{nunney2015peto}.

One of the most-discussed open problems in comparative cancer biology is due to Peto \cite{peto2016epidemiology} who famously remarked that, ``no plausible explanation has yet been offered for the fact that the risk of cancer in old age is not vastly different in species with very different life-spans''. He also noted here that humans have perhaps 1000 times as many cells as mice, with some risk of each cell becoming cancerous. Larger body size is correlated with longer lifespan across mammals, and both seemingly should predispose to greater cancer incidence. The apparent lack of this pattern is known as ``Peto's paradox''. The two issues of body mass and lifespan could in principle be addressed separately, but in practice have often been discussed together. In the literature on ``Peto's paradox'', some authors have focused on body mass, (e.g., \cite{noble2015peto}), while others have explicitly considered both body mass and lifespan, (e.g., \cite{tollis2017peto}). 

Several possible resolutions of Peto's paradox have been investigated. Caulin, Graham et al. \cite{caulin2015solutions} calculated that to reduce a whale's expected cancer risk to that of a human (with 1000-fold fewer cells) would require either a two- to three-fold decrease in the stem cell division rate, or two tumor-suppressor gene mutations. That paper compared mammalian genomes for differences in number of tumor-suppressor gene mutations. Although it reported evidence for genomic amplification of tumor-suppressor genes in several species, ranging in size from microbats to elephants, it did not find a positive correlation of tumor-suppressor genes with increasing body mass. 

In the absence of genomic explanations, physiology has received substantial attention. Comparison of human tissue types has shown that tissues with more tissue stem cell divisions have higher cancer risk \cite{noble2015peto, tomasetti2015variation}. This result further supports the suggestion that larger animals with a greater number of stem cell divisions would be expected to have increased cancer risk, and further demands some explanation for why this does not seem to be true. Absent systematic genetic differences, why is cancer risk not vastly higher for larger animals? 

Herman, Savage et al. \cite{herman2011quantitative} suggested that resolving Peto's paradox, may be possible by considering the scaling consequences of body mass and metabolic rate. Here we pursue this line of thinking, by including the related suggestion that cell-specific cancer risk is a function of cellular access to the resources allowing cell proliferation \cite{wu2019energy}. We therefore focus on scaling patterns for the cell-specific rate of vascular resource delivery. 

Allometric scaling of the cell-specific rate of energy use by somatic tissues has been explained by the observation that, ``generic properties of the vascular network constrain resource supply to cells.'' \cite{savage2007scaling}. In support of that explanation, empirical evidence indicates that somatic cells intrinsically tend to use energy substrates as rapidly as availability permits. Whereas rates of energy use by somatic cells in vivo are systematically lower in larger mammals, rates for cells in culture converge to a single, much higher, value for cells from all mammals regardless of body mass \cite{west2002allometric}. This has been interpreted as evidence that somatic-cell energy use and cell proliferation is extrinsically limited by the multicellular organism through control of the vascular delivery of proliferation-limiting energy substrates \cite{wu2019energy}. These authors suggested that limited vascular resource delivery is an organismal adaptation that limits somatic-cell proliferation and evolution, and thereby reduces cancer risk. Here we develop that idea by explicitly calculating rates of resource supply to cells across the full range of mammalian body size.

We begin with consideration of how waiting time to cancer scales with the supply of proliferation-limiting resources, based on reanalysis of unpublished quantitative simulation results of waiting time to cancer from \cite{wu2019energy} (Fig. 1). In order to understand cancer risk across diverse organisms and to resolve Peto's paradox, we combine these computational modeling results with consideration of how vascular delivery rates scale with body size, based on the network theory and ontogenetic growth framework of West, Brown et al. \cite{west2001general}.

For our analysis we focus on cancers of adulthood, even though rapid cell proliferation during development may explain much of childhood cancer and also contribute to cancers that appear in adults. We focus on adulthood both because adult cancers comprise most cases, and because they have been the focus of the literature on how cancer risk scales with body size. Our analysis considers cancer risk in adulthood in terms of lifetime totals and averages, and thus integrates certain carryover risks from childhood while still focusing on the cancers that appear in adulthood.

The theoretical work suggesting that vascular delivery is rate-limiting for somatic cell proliferation, and thus for oncogenesis, is consistent with substantial empirical evidence. Several different substrates can be rate-limiting for somatic cell proliferation. These include energetic substrates (e.g., glucose and oxygen), as well as crucial inorganic molecules such as phosphate \cite{bobko2017interstitial} and iron \cite{chen2019iron}. Each of these substrates are over-consumed by rapidly dividing neoplastic cells, making them scarce in neoplastic microenvironments. Each of these scarce and proliferation-limiting substrates is replenished only by vascular delivery. Therefore, vascular delivery is likely to be rate-limiting for cell proliferation, and therefore for oncogenesis and cancer progression \cite{herman2011quantitative, wu2019energy, savage2013using}. This is true even though tumors may differ in which specific substrate is rate-limiting for cell proliferation, because all proliferation-limiting resources arrive only through vascular delivery. These considerations make rates of cell-specific vascular delivery central to questions about cancer risk.

\section{Results}

\subsection{Effect of resource supply rate on cancer risk}

The explicit eco-evolutionary simulations of Wu, Aktipis et al. \cite{wu2019energy} indicated an inverse relationship between energy supply to tissues and waiting time for cancer, which serves as a proxy for cancer risk within a finite time period. These simulations were built on the assumption that in normal tissues without inflammation or wound healing, resting vascular delivery provides enough resources for somatic cells to function and maintain themselves, but not enough to proliferate beyond replacing lost cells. Under this assumption, limited vascular resource supply normally suppresses somatic cell evolution toward cancer, and any abnormal oversupply of resources accelerates somatic evolution, thereby increasing cancer risk. By fitting curves to those quantitative simulation results, we can calculate that the waiting time to cancer ($t_{c}$) in those simulation scales with resource delivery rate as: 
\begin{equation}
t_c\propto r^{-0.89}
\end{equation}
where $r$ is the multiple of resource delivery to tissues, and is defined as $r=R^{\prime}/R_{0}$ where $R_0$ and $R^{\prime}$ are the normal and elevated rates of resource delivery respectively. (see Appendices A3-A5, for technical details.) This implies that the waiting time for cancer can also be written as $t_c\propto R^{\prime -0.89}$. 

\subsection{Allometric scaling of cancer risk}

The results of the preceding section indicate that the waiting time to cancer depends on the cell-specific vascular delivery rate R, which we can compare across organisms of different body size. For mammals of varying body size, we can calculate R as the average adult total  cardiovascular output, $\langle Q \rangle$, divided by the average adult mass, $\langle M \rangle$, given that cell number N scales linearly with total body mass, $N\propto M^{1}$, \cite{savage2007scaling}. It should be noted that the total cardiovascular delivery rate is proportional to the total metabolic rate, $\langle Q \rangle\propto\langle B \rangle$ \cite{west2001general}, and we focus our derivations on the metabolic rate because it can be connected with the ontogenetic growth trajectories of West, Brown, et al. \cite{west2001general}. In Appendix A2, we use the ontogenetic growth model to calculate the quantities $B$ and $M$, where we find that $\langle B \rangle\propto\langle M \rangle^{3/4}$, and thus 
\begin{equation}
R=\langle B \rangle/\langle M \rangle\propto \langle M \rangle^{-1/4}.
\end{equation}
Thus, the cell-specific vascular delivery rate systematically decreases with increasing body size. If we define as a reference, the mass of the smallest mammal, $\langle M_0\rangle$, then the reference cell-specific rate of resource delivery, $R_0$, for that mass is
\begin{equation}
R_0=\langle B_0\rangle/\langle M_0\rangle\propto\langle M_0\rangle^{-1/4},
\end{equation}
from which the multiple of the reference cell-specific resource delivery for larger mammals is given by 
\begin{equation}
r\propto\langle M \rangle^{-1/4}/\langle M_0\rangle^{-1/4}. 
\end{equation}
Taken together these results combine to predict that the effect on oncogenesis of vascular delivery rate would influence time to cancer according to
\begin{equation}
t_c\propto\left(\langle M \rangle^{-1/4}\right)^{-0.89}\propto\langle M \rangle^{0.22}.
\end{equation}

A similar calculation of the waiting time to cancer (interpreted as the threshold time it takes for a given amount of energy to be delivered to a cell to stimulate cancerous growth) gives $t_c\propto\langle M \rangle^{1/4}$, which implies that $t_c\propto r^{-1}$, in approximate agreement with the result derived from the simulation of Wu et al. which gave an exponent of $\approx 0.89$.
	
All else being equal, these results show the increase in the waiting time to cancer for tissues of larger mammals because of slower resource delivery to cells. In isolation, this would suggest that larger mammals would have lower lifetime cancer risk owing to lower resource delivery rates to tissues. However, mammals have radically different lifespans and we need to take this into consideration to assess Peto's paradox. 

It is known that mammalian lifespans, $T$, scale as $T\propto\langle M \rangle^{0.21}$ \cite{speakman2005body}, which is very close to the scaling of time to cancer as $t_{c}\propto M^{0.22}$ (see above,also Appendix A4). This implies that total cancer risk over a lifetime would not vary systematically with body mass. Because waiting times to cancer and to normal end of life vary together with mass, the expectation of cancer arising before normal end of life does not change ad a function of mass. Our more detailed derivation in the Appendix also shows that the decreased cell-specific supply rate in larger mammals is counter-balanced by increased lifespans such that cancer risk per cell is expected to be approximately invariant across all mammalian body sizes. Figure 2A shows that at the expected lifespan the total energy delivered to a unit of tissue is the same across organisms of different size. Mammals reach this value at different rates according to lifespan. Another way to present this analysis is to look at the fraction of total energy delivered over a lifespan as a function of the fraction of a lifespan, which collapses all the lifespan curves for mammals onto a single universal curve (Figure 2B). The scaling of lifespan is the only normalization needed for this collapse.

\section{Discussion}

When we consider allometric scaling relationships, which were not part of the original framing of Peto's paradox, Peto's observations become less paradoxical. The apparent paradox arose from the assumption that large animals are like small animals in all respects, aside from having more cells and longer life spans, and that all physiological consequences including the risk of cancer naively increase accordingly. Available empirical data do not tell us quantitatively how the cell-specific risk of malignancy depends on access to proliferation-limiting resources. However, computer simulations of the process of somatic-cell evolution to malignancy have shown that this dependence does exist, and has a strong effect \cite{wu2019energy}. Thus, Peto's apparent paradox may be simply a result of the incorrect assumption of, ``all else equal'', with increasing body mass, including per-cell delivery of proliferation-limiting resources. In fact, the lower rate of resource delivery per cell in larger animals reduces cells' potential for proliferation and clonal evolution, and thus their risk of initiating cancer per unit time. However, this lower risk is compensated for by longer lifespans which match the waiting times for cancer. Our results indicate that the number of oncogenesis events per unit of body mass over a lifetime is expected to be roughly invariant across body mass. 

Peto (2016) addressed only differences between, and not within, species. Such allometric relationships have only been observed, and explained, between species. Therefore, we do not share the view of Nunney \cite{nunneyresolving} that changes in metabolic rates (or vascular delivery rates) should be discarded as explanations for Peto's paradox because this explanation is inconsistent with observed intraspecific effects of body size on cancer risk.

The usual functional interpretation of metabolic scaling is that restricted vascular delivery to body tissues of larger animals benefits the organism by reducing the energy costs of vascular delivery \cite{west1997general}. However, given the hypothesis of Wu, Aktipis et al. (2019) regarding the role of resource-dependent evolution of somatic cells, we suggest the observed allometric scaling of vascular delivery provides a second plausible benefit in the form of suppressed per-cell cancer risk in larger mammals. When life span exceeds waiting time to cancer, cancer incidence will be substantial, and its impacts on organismal fitness may drive selection for reduced vascular delivery per cell.

How might we explain the striking similarity between the scaling exponents of life span and of cellular waiting time to cancer? If we take as a given the evolved scaling of life span, and treat the scaling of vascular delivery as an adaptation to accommodate this evolved life span, this suggests selection to restrict vascular delivery to the point that waiting time to cancer approaches the typical life span for that body size, rendering cancer unimportant in reducing life span and fitness. Future work should focus on comparing the relative selective effects of hydrodynamic efficiency versus suppression of cancer risk.

It should be noted that our work addresses only the increased cancer risk to an individual associated with maintenance, and excludes the larger number of cell proliferations during development of larger animals. Future models could create a more complete picture of cancer risk by combining this developmental aspect with the risks examined here accompanying cell proliferation for tissue renewal.

\section{Figures}

\vspace*{-7pt}

\begin{figure}[h!]
\centering\includegraphics[width=0.5\textwidth]{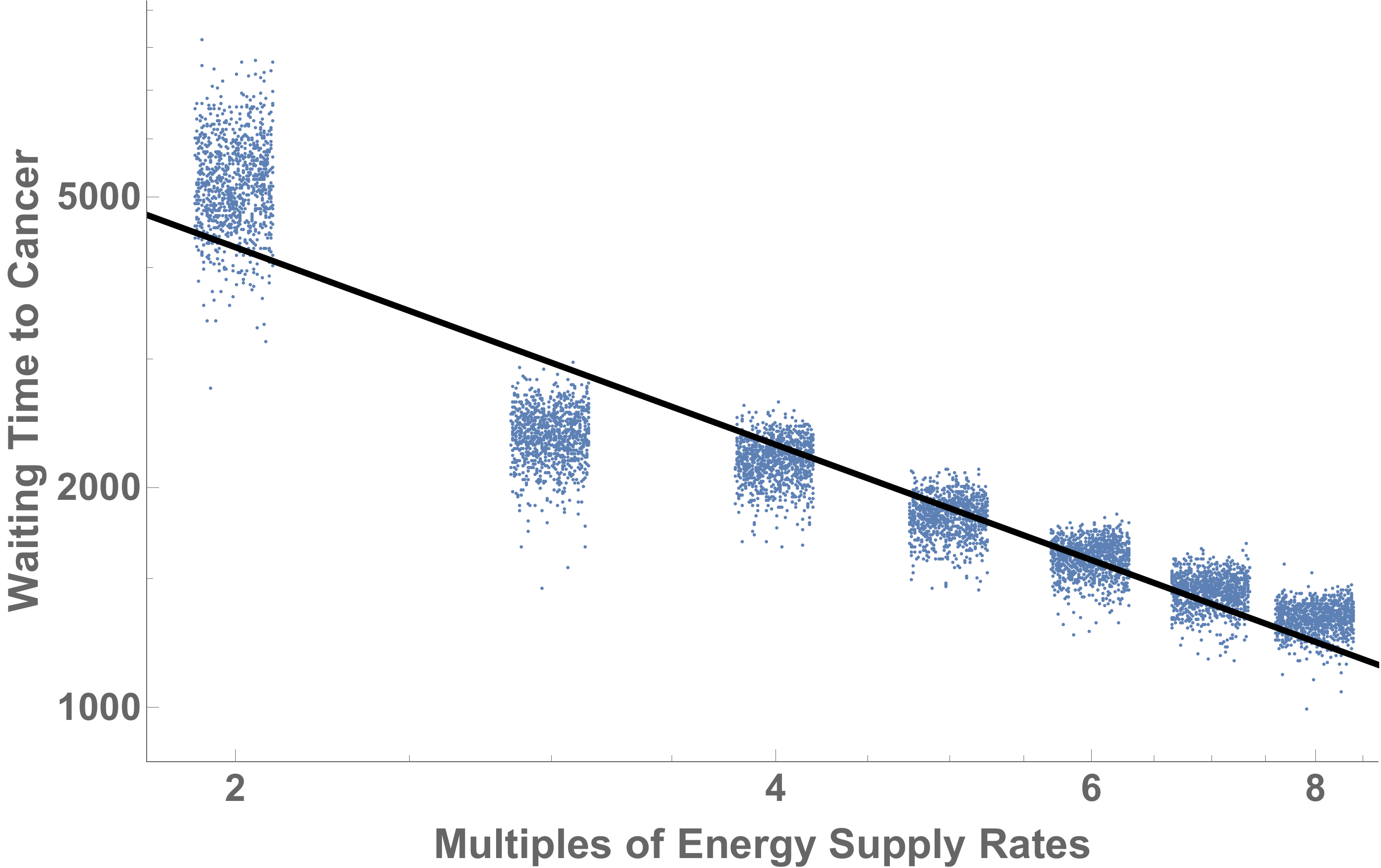}
\caption{The scaling of waiting time to cancer with energy supply rates, from computer simulations of cell evolution within tissues. The black line is the best fit to all the data, where the scaling exponent $\alpha=-0.89\pm0.01$. Each data point represents an independent evolutionary simulation where the x-axis values have an added random graphical factor to show all the data. The data are originally from the study reported by Wu, Aktipis, et al. \cite{wu2019energy}.}
\label{figure_1}
\end{figure}

\vspace*{-5pt}

\vspace*{-7pt}

\begin{figure}[h!]
\centering\includegraphics[width=1.0\textwidth]{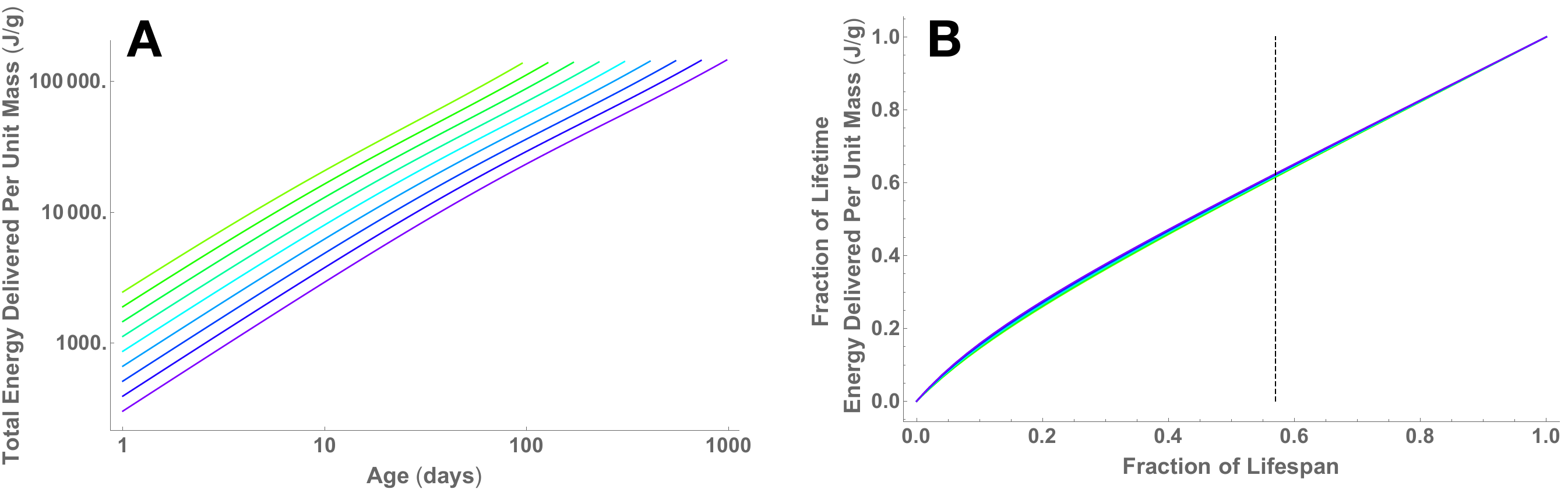}
\caption{(A) The total energy delivered to a unit of tissue as a function age and body size. The mammalian body sizes range from $100$ g (green curve) to $1,000,000$ g (purple curve). The total energy delivered per unit tissues reaches the same value at the expected lifespan for each body size. Note that the both axes have logarithmic scales, such that higher parallel lines have higher rates of increase with age, and larger mammals reach the same total energy delivered to tissues as smaller mammals, but do so at a slower rate. (B) The universal curve for percent of total lifetime energy delivered to each cell as a function of fraction of lifespan. The dashed line is the expected time to reach maturity.}
\label{figure_2}
\end{figure}

\vspace*{-5pt}

\vspace*{-7pt}

\begin{figure}[h!]
\centering\includegraphics[width=1.0\textwidth]{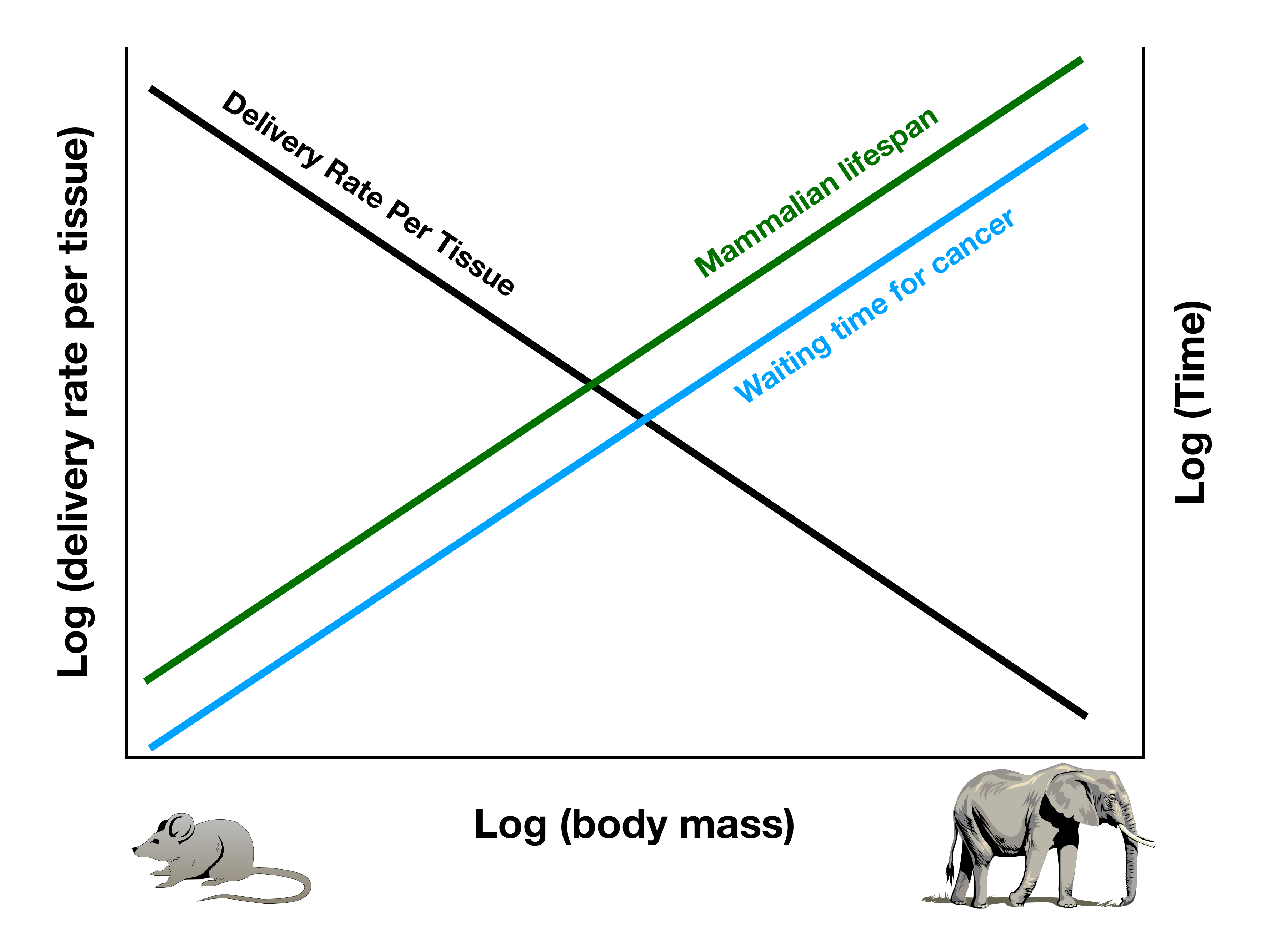}
\caption{The overall tradeoffs in resource delivery rates, lifespans, and waiting times for cancer as a function of mammalian body size that together lead to a constant cancer risk across all mammals.}
\label{figure_3}
\end{figure}

\vspace*{-5pt}

\newpage

\enlargethispage{20pt}


\dataccess{The data used in this paper are a reanalysis of ``Wu DJ, Aktipis A, Pepper JW. Energy oversupply to tissues: a single mechanism possibly underlying multiple cancer risk factors. Evolution, Medicine, and Public Health. 2019;2019(1):9-16.''}

\aucontribute{JWP, CPK, and GBW conceived of the study and discussed initial ideas for mathematical analysis. CPK and JWP carried out the mathematical analysis, produced figures, and wrote the initial manuscript draft. GBW provided feedback on the analyses contributed to the revision of the final manuscript.}

\competing{The authors declare no competing interests.}

\funding{CPK and GBW thank CAF Canada for generously supporting this work. JWP is employed by the National Cancer Institute.}

\ack{The authors also thank Van Savage, Alex Herman, and Eric Deeds for insightful conversations.}

\disclaimer{The opinions expressed by the author are their own and should not be interpreted as representing the official viewpoint of the National Cancer Institute, the National Institutes of Health or the U.S. Department of Health and Human Services.}



%

\section*{Appendices}

\setcounter{section}{0}
 \renewcommand*{\thesection}{A\arabic{section}}

\section{Lifetime Average Tissue Supply Rates}

Several previous efforts have derived growth rate from metabolic rate by considering that the total energy budget of an organism is partitioned between growth and maintenance purposes:
\begin{equation}
B_{0}m^{\alpha}=E_{m}\frac{dm}{dt}+B_{m}m
\label{budget}
\end{equation}
where $E_{m}$ is the energy required to synthesize a unit of new biomass, and $B_{m}$ is the metabolic rate required to maintain an existing unit of biomass. This equation can be solved for the growth trajectory of a single organism (e.g. \cite{west2001general,moses2008rmo,kempes2012growth}), and it can also be used for finding the growth rate of an organism. If we define $a=B_{0}/E_{m}$ and $b=B_{m}/E_{m}$, then we can rewrite equation \ref{budget} as 
\begin{equation}
\frac{dm}{dt}=am^{\alpha}-bm.
\label{diff-form}
\end{equation}
This equation can be solved \cite{kempes2012growth} to give the size trajectory of an organism
\begin{equation}
m\left(t\right)=\left[1-\left(1-\frac{b}{a}m_{0}^{1-\alpha}\right)e^{-b\left(1-\alpha\right)t}\right]^{1/\left(1-\alpha\right)}\left(\frac{a}{b}\right)^{1/\left(1-\alpha\right)},
\label{trajectory}
\end{equation}
where for mammals the metabolic scaling is given by $\alpha\approx3/4$. It should be noted that the asymptotic mass corresponds to $dm/dt=0$ in Equation \ref{diff-form}, leading to $M^{1-\alpha}=a/b$. This asymptotic mass is achieved only as time goes to infinity. Typically, $a$ is taken to be a constant as $B_{0}$ applies to an entire class of organisms conforming to one metabolic scaling relationship, and $E_{m}$ is found to be invariant across a variety of organisms. Thus, the maintenance term shifts across organisms of different adult size, because $b=a/M^{1-\alpha}=a/M^{1/4}$. Taken together this implies that $B_{m}\propto M^{-1/4}$ and that larger organisms spend less metabolic power on maintenance per unit mass during any portion of organism growth.  The above metabolic rates and ontogenetic growth curves allow us to calculate all relevant lifetime totals and averages for the argument presented in the main text and detailed below.

\section{Simple Lifetime Totals} 

Our argument rests on the time to cancer given the resource supply rate, $\left<R\right>$, to a unit of tissue. The simplest calculation considers the simple ratio of the lifetime average metabolic rate and organism mass. The lifetime average metabolic rate is given by 
\begin{equation}
\left<B\right>=\frac{1}{t_{death}}\int_{0}^{t_{death}}B_{0}m\left(t\right)^{3/4}dt
\label{lifetime-B}
\end{equation}
where $t_{death}$ is the lifespan of an organism. Similarly, the lifetime average mass is given by
\begin{equation}
\left<m\right>=\frac{1}{t_{death}}\int_{0}^{t_{death}}m\left(t\right)dt.
\label{lifetime-m}
\end{equation}
For these calculations it is essential to understand how the lifespan changes with organism size, which can be found by considering the time that it takes to reach a given fraction of the asymptotic mass in Equation \ref{trajectory} : $\epsilon M=m\left(t_{death}\right)$. For any fixed value of $\epsilon$ this relationship scales like 
\begin{equation}
t_{death}\propto M^{1/4}.
\label{gentime}
\end{equation}
This derived scaling of lifespan compares well with observations, where empirically it is known that $t_{death}=dM^{\beta}$ with $d=2.34$ (years g$^{-\beta}$) and $\beta=0.21$ \cite{speakman2005body,schmidt1984scaling,economos1980taxonomic}.  Using the scaling of lifetime it is possible to show that the solutions to Equations \ref{lifetime-B} and \ref{lifetime-m} are given by 
\begin{equation}
\left<B\right>\propto M^{3/4}
\end{equation} 
and
\begin{equation}
\left<m\right>\propto M.
\end{equation} 
Taken together these equations imply that
\begin{equation}
\left<B\right>\propto \left<m\right>^{3/4} 
\end{equation}
which is the main result needed for the calculations performed in the main text to find that $\left<R\right>\propto M^{-1/4}$. 

 \section{Detailed Lifetime Averages}

The most accurate assessment of average lifetime energy supply to tissues is given by considering resource supply rates at each point in ontogeny and averaging over these. This is given by 
\begin{eqnarray}
\left<R\right>&=&\frac{1}{t_{death}}\int_{0}^{t_{death}}\frac{B_{0}m\left(t\right)^{3/4}}{m\left(t\right)}dt\\
&=&\frac{1}{t_{death}}\int_{0}^{t_{death}}\frac{B_{0}}{m\left(t\right)^{1/4}}dt.
\label{ave-R}
\end{eqnarray}
This integral combined with the scaling for $t_{death}$ can be shown to produce 
\begin{equation}
\left<R\right>\propto M^{-1/4}
\end{equation} 
which is the same scaling result found from taking the ratio $\left<m\right>/\left<B\right>$. 


\section{Timescale Optimization for Constant Cancer Risk}

We can also turn these arguments around and ask what lifespan would result from all organisms dying of cancer. If cancer sets the timescale for death then we are interested in finding the point where $t_{death}=t_{c}$. The waiting time for cancer is given by 
\begin{equation}
t_{c}=cR^{\alpha}
\end{equation}
where $\alpha=-0.89$ (see an analysis of \cite{wu2019energy} below), and this allows us to find the appropriate timescale for death as
\begin{equation}
t_{death}=c\left<R\right>^{\alpha}=c\left[\frac{1}{t_{death}}\int_{0}^{t_{death}}\frac{B_{0}}{m\left(t\right)^{1/4}}dt\right]^{\alpha}.
\end{equation} 
Solving this equation for $t_{death}$ gives
\begin{equation}
t_{death}\propto M^{0.20}
\end{equation}  
which again scales similarly to known lifespans across mammals \cite{speakman2005body,schmidt1984scaling,economos1980taxonomic}. This result shows that one plausible argument for what sets lifespan scaling is the timescale for cancer formation. In this scenario the energetics associated with different body sizes leads to different times at which cancer forms and eventually kills an organism. It should be noted that this result still follows from the underlying scaling of metabolism with body size and all of the arguments for what sets that scaling (e.g. \cite{west1997general}).

\section{Estimating the scaling of cancer waiting times}

%
 
The explicit evolutionary simulations performed in \cite{wu2019energy} provide the connection between the latency to cancer and the energy supply rate. We reanalyzed the data to find the scaling relationships 
\begin{equation}
t_{c}\propto R^{\alpha}.
\end{equation}
The overall fit of this power-law to the data gives $\alpha=-0.89\pm0.01$. It should be noted that for low energy supply rates the data are slightly more noisy than for higher energy supply rates, and that the higher energy supply rates converge to a different power law characterized by $\alpha=-0.72\pm0.01$. For our analysis we used the value of $\alpha=-0.89\pm0.01$ found from fitting all of the data.

\end{document}